\documentstyle[12pt]{article}
\begin{document}
\def\applt{\;\; {\lower3pt\hbox{$
{\buildrel < \over {\scriptstyle \sim} }$}}\;\;}
\newcommand{\cl}{\centerline}
\def\d{{\rm d}}
\begin{titlepage}

\hfill{IP-ASTP-18-92}
\setlength{\textwidth}{5.0in}
\setlength{\textheight}{7.5in}
\setlength{\parskip}{0.0in}
\setlength{\baselineskip}{18.2pt}
\vfill
\cl{\large{{\bf  QCD Sum Rule and Perturbative QCD Approaches}}}\par
\cl{\large{{\bf  to Pion Compton Scattering}}}\par
\vskip 1.5cm
\cl{Claudio Corian\`{o}$^*$ and Hsiang-nan Li$^\dagger$}
\vskip 0.5cm
\cl{$^*$Institute for Theoretical Physics,
University of Stockholm}\par
\cl{S 113 Vanadisvagen, Stockholm, Sweden}\par
\cl{and}\par
\cl{Dipartimento di Fisica, Universita' degli Studi di Parma}\par
\cl{and INFN Gruppo Collegato di Parma, Italy}\par
\vskip 0.3cm
\cl{$^\dagger$Institute of Physics, Academia Sinica,}\par
\cl{Taipei, Taiwan 11529, R.O.C.}\vskip 1.0cm
\cl{\today}

\vskip 2.0 cm
\cl{\bf Abstract}

We compare two approaches to the description of pion Compton
scattering  at moderate momentum transfer, one being based
on local duality QCD sum rules for the invariant amplitudes of the process,
which have been derived recently,
and the other on the modified factorization
formula with Sudakov effects included. We find that perturbative QCD
predictions are dominant over those from QCD sum rules
only for the scattering angle greater than 90$^o$.

\vfill
\end{titlepage}

\noindent
{\em 1. Introduction}. The method of QCD sum rules  \cite{SVZ} has been
very successful in describing many of the static and dynamical properties
of hadrons at intermediate energy, a regime where non-perturbative
effects are important.
Based on this semi-phenomenological approach, important information on the
quark distribution functions of hadrons [2-4] and on their
electromagnetic form factors at momentum
transfer $Q$ about 1 to 2 GeV has been obtained [5-7].
On the other hand, the factorization formula, with Sudakov suppression taken
into account for soft gluon exchange,
has enlarged the applicability of perturbative QCD (pQCD) to
elastic hadron form factors down
to $Q\sim 2$-3 GeV, and produces reliable results in agreement with
experimental data \cite{LS}. The above energy
scales indicate the transition from non-pQCD to pQCD at $Q^2$ around 4 GeV$^2$.

Then, natural questions to ask are whether these two approaches
can be generalized to other more complicated processes, and on what energy
scale they give comparable predictions.
Recently, progress has been made
in extending the QCD sum rule analysis to a four-point amplitude in
pion Compton scattering \cite{CC,CR}. A crossed version of Compton
scattering, $p{\bar p}\rightarrow\gamma\gamma$, has been investigated based on
the modified factorization formula \cite{H}.
In this letter we will study pion Compton scattering following the
same approach, and compare the results with those from QCD sum rules.
Their behavior at moderate momentum transfer
may give hints about the above questions.
Besides, the additional angular dependence of
Compton scattering, which is absent in elastic form factors, will exhibit
more information on the transition to pQCD. Hence, Compton scattering
provides a nontrivial comparision between the two methods, and experiments
will determine which one gives a better description of the process.

In the sum rule approach, the lowest-order diagrams
for pion Compton scattering are those without virtual gluons,
as shown in fig.~1, which give real contributions \cite{CC,CR}.
The basic diagrams considered in pQCD
have one extra exchanged gluon as in fig.~2.
An imaginary contribution appears when
the exchanged gluons in fig.~2d and e (with two photons attaching to
different quark lines) go on-shell,
and thus brings in a phase \cite{FSZ}, which distinguishes the lowest-order
sum rule and pQCD approaches.
Since the singularity due to this on-shell
gluon is not removed by pion wave functions as in the case of form factors
\cite{MF}, Sudakov corrections are more important here in the sense of
improving the self-consistency of the perturbative formula.
Therefore, our work can be regarded as an
attempt to incorporate Sudakov effects into Compton scattering.

Since the main purpose is to compare the two methods, we do not
compute the full cross section of the process, but
concentrate on one of the invariant amplitudes contained in the
scattering matrix element. New sum rules for the invariant amplitude
have been derived in ref.~\cite{CR}.
A modified perturbative expression for the same invariant
amplitude will be given below.
We study the process in a ``brick-wall'' frame, where the incoming
pion momentum $p_1$ and outgoing one $p_2$ are assumed to have large
``$+$" and ``$-$" components respectively, and
${\bf p}_{T_1}={\bf p}_{T_2}=0$.
The incoming and outgoing photons, with momenta $q_1$ and $q_2$ respectively,
are on-shell $(q_1^2=q_2^2=0)$. The
Mandelstam invariants $s=(p_1+q_1)^2$, $t=(p_1-p_2)^2=-Q^2$ and
$u=(p_1-q_2)^2$ are restricted in the
physical region of $s>0$, $t<0$, and $u<0$.

To simplify the analysis, we assume that $s$ is larger
than $-t$, but not so large that higher-order corrections
proportional to $\ln(-t/s)$ diverge. It then suffices to consider
the contributions which are not suppressed by $s$ or $u$.
The ratio $-t/s$ characterizes the scattering
angle $\theta$ of the photon, $\sin^2(\theta/2)=-t/s$. The condition
$-t/s\rightarrow 0$ corresponds to forward
scattering of the photon, while $-t/s\rightarrow 1$ to backward scattering.
It is observed that the transition scale, on which the perturbative
contributions become comparable to non-perturbative contributions,
moves to large $-t$ as $\theta$ decreases. For forward scattering $(\theta
\rightarrow 0)$ non-perturbative contributions always dominate.
Only for $-t/s > 0.5$ $(\theta > 90^o)$ are pQCD predictions important
at moderate momentum transfer.

\vskip 0.5cm
{\em 2. QCD Sum Rule Approach}.
A detailed QCD sum rule analysis of pion Compton scattering
has been developed in \cite{CC,CR}. In this section we will
emphasize the
essential aspects of the analysis and quote the results.
We start with the correlator of 4-currents
\begin{eqnarray}
\Gamma_{\alpha\mu\nu\beta}(p_1^2,p_2^2,s,t)&=&i\int{\rm d}^4x\,{\rm d}^4y
\,{\rm d}^4z\exp
(-ip_1\cdot x+ip_2\cdot y-iq_1\cdot z)
\nonumber \\
& &\times \langle 0|T\left(\eta_{\beta}(y)J_{\nu}(0)J_{\mu}(z)
\eta_{\alpha}^{\dagger}(x)\right)|0\rangle \; ,
\label{tp}
\end{eqnarray}
where
\begin{eqnarray}
J_{\mu}=e_u\bar{u}\gamma_{\mu}u+e_d\bar{d}\gamma_{\mu}d,
\;\;\;\;\;\;
\eta_{\alpha}=
\bar{u}\gamma_5\gamma_{\alpha}d
\label{jd}
\end{eqnarray}
are the electromagnetic and
axial currents respectively in terms of quark fields $u$, $d$ and
charges $e_u=2/3$, $e_d=-1/3$. The pions are assumed to be off-shell
$p_1^2=s_1\not= 0$, $p_2^2=s_2\not=0$ with $s_1$ and $s_2$ their
virtualities.

$\Gamma_{\alpha\mu\nu\beta}$ can be expressed in terms of many structures,
each of which is associated with an invariant amplitude.
We study the invariant ampltudes $H_1$ and $H_2$ with the structure
proportional to $p_{1\alpha}p_{2\beta}$ \cite{CR}
\begin{equation}
\Gamma_{\alpha\mu\nu\beta}\propto f_{\pi}^2p_{1\alpha}p_{2\beta}
\left(H_1e^{(1)}_{\mu}e^{(1)}_{\nu}+H_2e^{(2)}_{\mu}e^{(2)}_{\nu}\right)+...\;
,
\end{equation}
where the helicity vectors $e^{(1)}$ and $e^{(2)}$ are defined by
\begin{eqnarray}
& &e^{(1)\mu}=\frac{N^{\mu}}{\sqrt{-N^2}}\;,\;\;\;\;\;\;
e^{(2)\mu}=\frac{P^{\mu}}{\sqrt{-P^2}}
\label{e1}
\end{eqnarray}
with
\begin{eqnarray}
& &N^{\mu}=\epsilon^{\mu\lambda\sigma\rho}P_{\lambda}q_{\sigma}K_{\rho}\;,
\nonumber \\
& &P^{\mu}=p_1^{\mu}+\nu p_2^{\mu}-K^{\mu}
\frac{p_1\cdot K+\nu p_2\cdot K}{K^2}\; ,
\nonumber \\
& &\nu={p_1\cdot p_2 -s_1\over p_1\cdot p_2 - s_2}\;,  \nonumber \\
& &K=q_1+q_2\;,\;\;\;\;\;\;q=q_2-q_1\;.
\end{eqnarray}
The structure $p_{1\alpha}p_{2\beta}$ comes from the insertion of pion
states into the correlator in eq.~(\ref{tp}) and the substitution
\begin{eqnarray}
& &\langle \pi(p_1)|\eta_{\alpha}^{\dagger}(0)|0\rangle =
-if_{\pi}p_{1\alpha}\;, \nonumber \\
& &\langle 0|\eta_{\beta}(0)|\pi(p_2)\rangle =if_{\pi}p_{2\beta}\;,
\end{eqnarray}
which interpolates between the vacuum and single pion state.
We concentrate on the combination $H=H_1+H_2$, which is extracted by
contracting
$\Gamma_{\alpha\mu\nu\beta}$ with $-g^{\mu\nu}n^{\alpha}n^{\beta}$, where
the circular polarization
\begin{equation}
n^\mu= (e^{(2)} \pm i e^{(1)})^\mu
\end{equation}
satisfies the conditions
\begin{equation}
n\cdot q_1=n\cdot(p_1-p_2)=n^2=0\;.
\label{n}
\end{equation}
These conditions, in full analogy with those in the case of form factors
\cite{NR1}, are crucial for obtaining the correct asymptotic
behavior of the sum rules, which is consistent
with quark counting rules \cite{seventeen,eighteen}.
Note that the expression for $e^{(2)}$
given in eq.~(\ref{e1}) is slightly different from that in
\cite{LL}; the vector $P^\mu$ is defined
including dependence on the extra factor $\nu$ due to the off-shell pions.
This modified definition approaches the standard one
given by $\nu=1$ at the pion pole, and
guarantees the existence of $n^\mu$ with the
requirements of eq.~(\ref{n}).

To evaluate the helicity form factor $H(s,t)$, it is simplest to
consider its spectral function
$\Delta(s_1,s_2,s,t)$ which is defined by  the double discontinuity of
$H$ on the cuts $0\leq s_1,\;s_2\leq \lambda^2 \approx (s+t)/4$.
This choice of $\lambda$ is to avoid the inclusion of $u$-channel
resonances. The sum rule for $H(s,t)$, which is closely related
to the finite energy sum rule in \cite{LST},
has been formulated using the argument of analyticity for
the correlator in eq.~(\ref{tp}) in the finite region of the $s_1$ and $s_2$
complex planes \cite{CR}. The explicit form of $\Delta$ is essential both in
the derivation of the sum rule
and in the calculation of power corrections \cite{CL}.
 We will evaluate $\Delta$
using the standard cutting rules as developed in \cite{JS}.
It is impossible to render all internal lines on-shell
because of the off-shell assumption for the external pions.
Only the case with all lines on-shell except the upper line contributes.
We refer to \cite{CR} for a comprehensive
derivation for the sum rule, and quote the final expression here
\begin{eqnarray}
f_{\pi}^2H(s,t)&=&\int_0^{s_0}{\rm d}s_1
\int_0^{s_0}{\rm d}s_2\,\rho(s_1,s_2,s,t)
\exp(-\frac{s_1+s_2}{M^2})
\nonumber \\
& &+c_1\frac{\langle 0|G_{\mu\nu}G^{\mu\nu}|0\rangle}{M^4}
+c_2\frac{\langle 0|({\bar q}q)^2|0\rangle}{M^6}\; ,
\label{bt}
\end{eqnarray}
$f_{\pi}=133$ MeV being the pion decay constant.
The spectral density $\rho$ is given by
\begin{eqnarray}
\rho(s_1,s_2,s,t)&=&{1\over (n\cdot p_1)(n\cdot p_2)}\Delta(s_1,s_2,s,t)
\nonumber \\
&\approx&\frac{10}{3(2\pi)^2\delta^3}[(s_1-s_2)^2-t(s_1+s_2)]\left(
\frac{2s-\delta}{s}+\frac{2u-\delta}{u}\right)\; ,
\nonumber \\
\delta&=&\sqrt{(s_1+s_2-t)^2-4s_1s_2}\;.
\end{eqnarray}
Those terms suppressed by extra powers of $t$ or $s$ have been neglected.
The gluon and quark condensates $\langle 0|G_{\mu\nu}G^{\mu\nu}|0\rangle$
and $\langle 0|({\bar q}q)^2|0\rangle$ , arising from
the operator product expansion of the correlator
in eq.~(\ref{tp}), give the power corrections which are suppressed
by factors $(1/M^2)^n$, $M^2$ being the Borel mass.
The coefficients $c_1$ and $c_2$
can be obtained through the operator product expansion.
The exact value of the quark-hadron duality interval
$s_0$ is determined by demanding that the right-hand side of eq.~(\ref{bt})
have the weakest sensitivity to the variation of $M^2$ [2-6].
Once $s_0$ is specified, the large $M^2$ limit is applied to diminish
the power corrections, and the final
expression for $H$ is written as a local duality approximation
\begin{eqnarray}
f_{\pi}^2H(s,t)=\int_0^{s_0}{\rm d}s_1
\int_0^{s_0}{\rm d}s_2\,\rho(s_1,s_2,s,t)\;,
\label{bt1}
\end{eqnarray}
which behaves like $1/Q^4$ asymptotically.
We will address the calculation of the power corrections in a seperate
work \cite{CL}, and assign an approximate value to $s_0$ here.
It is also possible to derive a similar sum rule for the single invariant
amplitude $H_1$.

\vskip 0.5cm
{\em 3. Perturbative QCD approach}. Part of
the basic diagrams for the perturbative QCD approach are shown in fig.~2.
Each represents a class of diagrams which can be transformed to each
other by permuting the incoming and outgoing
pions, or the two quark lines. They differ from fig.~1
by an extra exchanged gluon. We will incorporate Sudakov effects for soft
gluon exchange, the case in which the running coupling constant $\alpha_s$
with its argument set to the gluon energy is large and lowest-order
pQCD does not make sense.  The method to calculate pion Compton
scattering based on these diagrams
is similar to that developed for electromagnetic form factors \cite{LS}.
Following the same reasoning and procedures, the modified
factorization formula for $H(s,t)$ is given by
\begin{eqnarray}
H(s,t)&=&\sum_{l=1}^2\int_0^1\; \d x_{1}\d x_{2}
\d^2{\bf k}_{T_1}\d^2{\bf k}_{T_2}\;
\psi(x_2,{\bf k}_{T_2},p_{2}) \nonumber \\
& &\times
T_{H_l}(x_i,s,t,{\bf k}_{T_i})\,
\psi(x_1,{\bf k}_{T_1},p_{1})\; .
\label{2}
\end{eqnarray}
The additional dependence on transverse momentum ${\bf k}_T$ carried
by a valence quark has been included into the pion wave function
$\psi$ and the hard-scattering kernel $T_H$. Eq.~(\ref{2}) can be understood
as an intermediate step in the standard factorization program \cite{BS},
where ${\bf k}_T$ in $T_H$ is assumed to give higher-power
$({\bf k}_T^2/Q^2)$ correction and thus ignored. In fact, this approximation
is not proper when the exchanged gluon becomes soft.
The contributions to the hard scatering
from each diagram in fig.~2, obtained by contracting the two photon vertices
with $-g^{\mu\nu}$, are given in table.~1.
The transverse momenta on the virtual
quark lines have been neglected in the derivation
since they are associated with linear,
instead of quadratic, divergences in $x_i$, and hence less important
than that on the gluon line \cite{LS}.
The contributions from fig.~2d and e, like fig.~1c, are suppressed by $s$
as stated in the previous section. All contributions
are grouped into only two terms $(l=1,\;2)$ using the
permutative symmetry.

Rewriting eq.~(\ref{2}) in terms of the Fourier transformed functions,
and inserting
the large-$b$ asymptotic behavior of the wave function \cite{BS}, we have
\begin{eqnarray}
H(s,t)&=& \sum_{l=1}^2 \int_{0}^{1}\d x_{1}\d x_{2}\,
\phi(x_{1})\phi(x_{2})
\int_{0}^{\infty} b\d b
{\tilde T}_{H_l}(x_i,s,t,b,w_l)
\nonumber \\
& & \times \exp[-S(x_i,b,Q,w_l)]\; .
\label{15}
\end{eqnarray}
where $b$, introduced by
Fourier transformation, is the seperation between the two valence quark lines.
Note the extra Sudakov factor $\exp(-S)$ compared to the standard
factorization formula, which arises from an all-order
summation of the collinear
enhancements of radiative corrections to fig.~2.
The exponent $S$ is written as \cite{LS}
\begin{equation}
S(x_{1},x_{2},b,Q,w)=\sum_{i=1}^{2}(s(x_{i},b,Q)+s(1-x_{i},b,Q))-
\frac{2}{\beta_{1}}{\rm ln}\frac{\hat{w}}{-\hat{b}}\; ,
\label{16}
\end{equation}
with
\begin{eqnarray}
s(\xi,b,Q)&=&\frac{A^{(1)}}{2\beta_{1}}\hat{q}\ln\left(\frac{\hat{q}}
{-\hat{b}}\right)+
\frac{A^{(2)}}{4\beta_{1}^{2}}\left(\frac{\hat{q}}{-\hat{b}}-1\right)-
\frac{A^{(1)}}{2\beta_{1}}(\hat{q}+\hat{b})
\nonumber \\
& &-\frac{A^{(1)}\beta_{2}}{16\beta_{1}^{3}}\hat{q}\left[\frac{\ln(-2\hat{b})
+1}{-\hat{b}}-\frac{\ln(2\hat{q})+1}{\hat{q}}\right]
\nonumber \\
& &-\left[\frac{A^{(2)}}{4\beta_{1}^{2}}-\frac{A^{(1)}}{4\beta_{1}}
\ln\left(\frac{e^{2\gamma-1}}{2}\right)\right]
\ln\left(\frac{\hat{q}}{-\hat{b}}\right)
\nonumber \\
& &-\frac{A^{(1)}\beta_{2}}{32\beta_{1}^{3}}\left[
\ln^{2}(2\hat{q})-\ln^{2}(-2\hat{b})\right]\; .
\end{eqnarray}
The variables ${\hat q}$, ${\hat b}$ and ${\hat w}$ are defined by
\begin{eqnarray}
{\hat q} &\equiv & {\rm ln}\left [\xi Q/(\sqrt{2}\Lambda)\right ]
\nonumber \\
{\hat b} &\equiv & {\rm ln}(b\Lambda)
\nonumber \\
{\hat w} &\equiv & {\rm ln}(w/\Lambda)\; ,
\label{11}
\end{eqnarray}
where the scale parameter $\Lambda\equiv
\Lambda_{\rm QCD}$ will be set to 0.1 GeV.
The coefficients $\beta_i$ and $A^{(i)}$ are
\begin{eqnarray}
& &\beta_{1}=\frac{33-2n_{f}}{12}\;,\;\;\;\;\beta_{2}=\frac{153-19n_{f}}{24}\;
,
\nonumber \\
& &A^{(1)}=\frac{4}{3}\;,
\;\;\;\; A^{(2)}=\frac{67}{9}-\frac{\pi^{2}}{3}-\frac{10}{27}n_
{f}+\frac{8}{3}\beta_{1}\ln\left(\frac{e^{\gamma}}{2}\right)
\label{12}
\end{eqnarray}
with $n_f=3$ the number of quark flavors and $\gamma$ the Euler constant.
The Sudakov factor is always  less than 1 as explained in ref.~\cite{LS},
and decreases quickly in the large-$b$ region.
The function $\phi$, obtained by factoring the $Q$ and $b$ dependences
from the transformed wave function
into Sudakov logarithms, is
taken as the Chernyak and Zhitnitsky model \cite{CZ1}
\begin{equation}
\phi^{CZ}(x)=\frac{15f_{\pi}}{\sqrt{2N_{c}}}\,x(1-x)(1-2x)^{2}\; ,
\end{equation}
where $N_{c}=3$ is the number of colors.

The transformed hard scatterings ${\tilde T}_{H_l}$
are given by
\begin{eqnarray}
{\tilde T}_{H_1}
&=&\frac{16\pi{\cal C}_F(e_u^2+e_d^2)\alpha_s(w_1)}{(1-x_1)(1-x_2)}
K_0\left(\sqrt{|r_1|}b\right)\left(\frac{[(1-x_1)t+u][(1-x_2)t+u]}{s^2}
\right.
\nonumber \\
& &\left.+\frac{[(1-x_1)t+s][(1-x_2)t+s]}{u^2}-4(1-x_2)\right)
\label{h1}
\end{eqnarray}
from the classes of fig.~1a-c, and
\begin{eqnarray}
{\tilde T}_{H_2}&=&32\pi{\cal C}_F e_u e_d\alpha_s(w_2)\left[\theta(-r_2)
K_0\left(\sqrt{|r_2|}b\right)
-\theta(r_2)\frac{i\pi}{2}H_0^{(1)}\left(\sqrt{r_2}b\right)\right]
\nonumber \\
&  & \times\left(\frac{1}{x_1(1-x_1)}-
\frac{(1+x_2-x_1 x_2)t^2+(1+x_2-x_1)ut}{x_2(1-x_{1})s^2}\right.
\nonumber \\
& &\;\;\;\;\left.+\frac{1}{x_2(1-x_2)}-
\frac{(1+x_1-x_1x_2)t^2+(1+x_1-x_2)st}{x_1(1-x_{2})u^2}\right)
\label{hh}
\end{eqnarray}
from the class fig.~1d-e with
\begin{eqnarray}
r_1=x_1x_2t,\;\;\;\;\;\;r_2=x_{1}x_{2}t+x_1u+x_2s\; .
\end{eqnarray}
$K_{0}$ and $H_0^{(1)}$ in eqs.~(\ref{h1}) and (\ref{hh}) are the
Bessel functions in the standard notation. The imaginary contribution comes
from eq.~(\ref{hh}).
The argument $w_l$ of $\alpha_s$
is defined by the largest mass scale in the hard scattering,
\begin{equation}
w_1=\max\left(\sqrt{|r_1|},\frac{1}{b}\right),\;\;\;\;\;\;
w_2=\max\left(\sqrt{|r_2|},\frac{1}{b}\right)\; .
\label{9}
\end{equation}
As long as $b$ is small, soft $r_l$  does not lead to large $\alpha_s$.
Therefore, the non-perturbative region in the modified factorization
is characterized by large-$b$, the region which is suppressed by Sudakov
effects. Eq.~(\ref{15}), as a perturbative expression, is thus
relatively self-consistent compared to
the leading-power factorization \cite{KN}.

\vskip 0.5cm

{\em 4. Numerical Analysis}.
To evaluate the invariant amplitude $H(s,t)$ from the sum rule,
 the value of $s_0$ must be determined
through a stability analysis
on the $M^2$ dependence of eq.~(\ref{bt}) as in [2-6].
However, we know that $s_0$
must take a value between the masses of pion $(m_{\pi}^2=0)$ and A$_1$-meson
$(m_{A_1}^2=1.2\;\;{\rm GeV}^2)$. Hence,
we estimate that $s_0$ resides in the reasonable
range $0.6<s_0< 0.8$ GeV$^2$, as referred to the case of pion form factor.
$H(s,t)$ with $s_0=0.6$, 0.7 and 0.8 GeV$^2$
are computed and found to vary slowly in
the above range of $s_0$ for $Q^2 > 4$ GeV$^2$. Therefore, $s_0$ will be set
to 0.7 GeV$^2$ in eq.~(\ref{bt1}) for the sum rule analysis. Results of
$Q^2H$ for different $-t/s$
are shown in fig.~3 with $4< Q^2< 25$ GeV$^2$.
It is observed that $H$ decreases
monotonously following the
$1/Q^4$ asymptotical behavior, and has a weak angular dependence.

Results derived from the modified expression
eq.~(\ref{15}) in the same range of $Q^2$ are also
shown in fig.~3, where $|H|$ denotes the magnitude of $H$. It is found that
$|H|$ has only little (logarithmical) deviation from the
expected  $1/Q^2$ behavior, and drops quickly with decreasing $-t/s$.
That is, pQCD predictions have a stronger angular dependence.
The curve from eq.(\ref{15}) denoted by $-t/s=0.6$ $(\theta=100^o)$
is always above
the corresponding curve derived from the sum rule for $4<Q^2<25$ GeV$^2$.
It indicates the dominance of perturbative contribution in large-angle
Compton scattering.
For $-t/s=0.5$ $(\theta=90^o)$ sum rule and pQCD predictions are comparable,
and the transition scale is about 6 GeV$^2$.
As $-t/s$ drops, the transition scale increases, and the non-perturbative
contribution becomes dominant when $-t/s$ goes down to 0.2 $(\theta=50^o)$.
Based on the above analysis, we conjecture that
the present pQCD calculation for proton Compton scattering \cite{KN}
or its crossed version like $p{\bar p}\rightarrow \gamma\gamma$ \cite{H}
should be
complemented by the QCD sum rule description at small angles.

The phase $\phi$ of $H$ obtained from eq.~(\ref{15})
is also studied, and its $Q^2$ dependence for different $-t/s$
is shown in fig.~4. It is found that $\phi$ is very stable at intermediate
angles, but changes drastically at small angles.

The corresponding results from the standard factorization formula
with ${\bf k}_T$ neglected in $T_H$ and $\alpha_s= 0.3$
in eq.~(\ref{2}) are exhibited
for reference, and are similar to those from the modified one.
This choice of $\alpha_s$ has been used in order to fit experimental data
from proton Compton scattering \cite{KN}. However,
$|H|$ shows an exact $1/Q^2$ behavior and $\phi$ is independent
of $Q^2$ at a fixed angle. Note the difference between the standard and
modified factorizations that
$\alpha_s$ is effectively regarded as a free parameter in the former case.

\vskip 0.5cm

{\em 5. Conclusion}.
In this letter we have shown that the extension of both the QCD
sum rule method and modified factorization formula to pion Compton
scattering is possible, because their agreement
at intermediate scattering angles is explicit. We believe that
this coincidence is
not trivial due to the very different theoretical bases of the two
approaches.  The transition scale from QCD sum rules to pQCD
is found to vary with scattering angles. QCD sum
rules give important contribution at small angles, and
pQCD becomes dominant only for
$\theta > 90^o$ from our analysis. Our results have suggested
a new and interesting
interplay between factorization theorems and sum rule methods in Compton
scattering. All these observations need to be
compared with experiments.
To examine the conclusions more exactly, it is
necessary to determine a precise value for $s_0$,
and thus the calculation for gluon and quark condensates, along with a
detailed discussion on the stability of the sum rule, is
involved \cite{CL}.
It is also of interest to
generalize the two methods to more complicated case like proton
Compton scattering \cite{KN}, for which experimental data are available.

\vskip 0.5cm
We thank Prof. G. Sterman for suggesting this investigation,
and  Profs. T.H. Hansson and H. Rubinstein
for clarifying discussions. C. C. thanks Prof. G. Marchesini
and Dr. F. Fiorani for stimulating conversations and for their hospitality
at the Physics Department of Parma University, and Profs. N. Isgur and
A.V. Radyushkin for their hospitality at CEBAF.
This work was supported in part by the National
Science Council of R.O.C. under Grant No. NSC82-0112-C001-017 and
by INFN of Italy.

\newpage
Table 1. The expressions of the hard scattering $T_H$
corresponding to the diagrams in fig.~2.

\[ \begin{array}{rc}   \hline\hline\\
{\rm Diagram} & T_H/(64\pi\alpha_{s}) \\
\hline   \\
({\rm a})     &{\displaystyle
\frac{-e_{u}^2[(1-x_1)t+u][(1-x_2)t+u]}
{(1-x_1)(1-x_2)s^2[x_1x_2t-({\bf k}_{T_1}
-{\bf k}_{T_2})^{2}]}}\\
        &    \\
({\rm b})   &  {\displaystyle
\frac{e_{u}^2}
{(1-x_{1})[x_{1}x_2t-({\bf k}_{T_1}
-{\bf k}_{T_2})^{2}]}}\\
        &   \\
({\rm c})   &  {\displaystyle
\frac{e_{u}^2}
{(1-x_{1})[x_{1}x_2t-({\bf k}_{T_1}
-{\bf k}_{T_2})^{2}]}}\\
        &   \\
({\rm d})   &  {\displaystyle
\frac{-e_{u}e_d}
{x_1(1-x_{1})[x_{1}x_{2}t+x_1u+x_2s-({\bf k}_{T_1}-{\bf k}_{T_2})^{2}]}}\\
        &    \\
({\rm e})   & {\displaystyle
\frac{e_{u}e_d[(1+x_2-x_1x_2)t^2+(1+x_2-x_1)ut]}
{x_2(1-x_{1})s^2[x_{1}x_{2}t+x_1u+x_2s-({\bf k}_{T_1}-{\bf k}_{T_2})^{2}]}}
\vspace{0.2cm}\\
\hline\hline
\end{array}   \]

\newpage
\cl{\large \bf Figure Captions}
\vskip 0.5cm

\noindent
{\bf Fig. 1.} Diagrams for the pion Comptom scattering in the QCD sum
rule analysis.
\vskip 0.5cm

\noindent
{\bf Fig. 2.} Part of the diagrams for the pion Comptom scattering
in the perturbative QCD analysis.
\vskip 0.5cm

\noindent
{\bf Fig. 3.} Dependence of $Q^{2}|H(s,t)|$ on $Q^2$ derived from QCD sum
rules (dashed lines), from the modified factorization (solid lines) and from
the leading-power factorization (dotted lines)
for (a) $-t/s=0.6$ $(\theta=100^o)$, (b) $-t/s=0.5$ $(\theta=90^o)$
and (c) $-t/s=0.2$ $(\theta=50^o)$.
\vskip 0.5cm

\noindent
{\bf Fig. 4.} Dependence of the phase $\phi$ on $Q^2$ from the
modified factorization (solid lines) and from
the leading-power factorization (dotted lines)
for (a) $-t/s=0.6$, (b)
$-t/s=0.5$, (c) $-t/s=0.2$ and (d) $-t/s=0.02$.

\end{document}